\documentclass[oneside]{article}
\usepackage[T1]{fontenc}
\usepackage{authblk}
\usepackage{float}
\usepackage{amsmath}
\usepackage{graphicx}
\usepackage{xfrac}
\usepackage[english]{babel}
\usepackage{lmodern}
\usepackage[T1]{fontenc}

\usepackage{csquotes}

\usepackage{longtable}

\usepackage{mathtools}
\usepackage{graphicx}
\usepackage{esint}
\usepackage[unicode=true,
 bookmarks=false,
breaklinks=false,pdfborder={0 0 1},backref=false,colorlinks=false]
 {hyperref}
\usepackage[a4paper, total={210mm,297mm},margin=2.0cm]{geometry}

\usepackage{upgreek}

\title{Multiscale Hybrid Modeling of Proteins in Solvent: SARS-CoV2 Spike Protein as test case for Lattice Boltzmann - All Atom Molecular Dynamics Coupling}
\author[1]{M. Lauricella}
\author[2]{L. Chiodo\thanks{Electronic address: \texttt{m.lauricella@iac.cnr.it}; Corresponding author}}
\author[3]{F. Bonaccorso}
\author[4]{M. Durve}
\author[5]{A. Montessori}
\author[1]{A. Tiribocchi}
\author[2]{A. Loppini}
\author[2]{S. Filippi}
\author[4]{S. Succi}

\affil[1]{Consiglio Nazionale delle Ricerche, Istituto per le Applicazioni del Calcolo IAC-CNR, 00185, Rome Italy.}
\affil[2]{Engineering Department, Campus Bio-Medico University, 00128, Rome, Italy.}
\affil[3]{Physics Department and INFN, University of Rome “Tor Vergata”, 00133 Rome, Italy.}
\affil[4]{Istituto Italiano di Tecnologia, Center for Life Nano- \& Neuro-Science@Sapienza – IIT, 00161, Rome, Italy.}
\affil[5]{Engineering Department, Universit\'{a} degli Studi Roma TRE, via Vito Volterra 62, Rome, 00146, Italy.}

\date{\today}

\begin{document}

\maketitle

\begin{abstract}
Physiological solvent flows surround biological structures triggering therein collective motions. Notable examples are virus/host-cell interactions and solvent-mediated allosteric regulation. The present work describes a multiscale approach joining the Lattice Boltzmann fluid dynamics (for solvent flows) with the all-atom atomistic molecular dynamics (for proteins) to model functional interactions between flows and molecules. 
We present, as an applicative scenario, the study of the SARS-CoV-2 virus spike glycoprotein protein interacting with the surrounding solvent, modeled as a mesoscopic fluid. The equilibrium properties of the wild-type spike and of the Alpha variant in implicit solvent are described by suitable observables. The mesoscopic solvent description is critically compared to the all-atom solvent model, to quantify the advantages and limitations of the mesoscopic fluid description.  
\end{abstract}

\section{Introduction}
\label{sec1}
In the last decades, continuous methodological and technological progress paved the successful path of computational tools in fighting contagious diseases by providing in silico simulations of biological molecules and drug design. Further, the graphics processing units (GPUs) granted significant technological progress, boosting the computational power over large system sizes. In this context, we use a multiscale modeling approach that couples a mesoscale solvent representation to the molecular dynamics. The final aim is to deliver an efficient biophysical computational  strategy to boost the simulation power and understand the biological mechanisms at the atomistic level, avoiding the computational effort due to the atomistic description of the solvent. 

Nowadays, the molecular dynamics (MD) method has shown its massive potential in characterizing the biological mechanisms underlying the activities of several proteins at the atomistic level. Remarkable examples in computational biophysics are the recent simulations of an entire cell organelle, a photosynthetic chromatophore vesicle from a purple bacterium \cite{rochaix2019dynamic} or the study of the N-Methyl-D-Aspartate (NMDA) neuroreceptor by the DE Shaw research group \cite{song2018mechanism}. As of today, an enormous scientific effort has been spent to investigate in-silico the molecular behavior of SARS-CoV-2 proteins, both for drug repurposing and for antibody design \cite{gadioli2021exscalate}. Standard MD simulations have been used, for example, to estimate binding free energies of spike in interaction with the human angiotensin-converting enzyme 2 (ACE2) receptor \cite{taka2021critical,he2020molecular,armijos2020sars,zou2020computational} alongside with their interaction scores \cite{brielle2020sars}. 
The effort devoted to SARS-CoV-2 proteins, with the exceptional focus on its spike protein, somehow put the spotlight on the strengths and limitations of bioinformatics and biophysics computational tools\cite{gao2021methodology,muratov2021critical} in the field of medicine and drug discovery, bringing these tools also to the attention of the general public. 

The widely known and investigated spike protein is here used as a test case to highlight strengths and drawbacks of a mixed multiscale description scheme. 
Indeed, despite the all-atom molecular dynamics description being the method of election to properly describe the biochemical nature of protein functioning, one of the main issues in its usage is related to the long time scales of biological mechanisms, and on their rare-event nature from a statistical mechanics perspective. 
Normally, the quaternary movements associated with the allosteric and functional response of biological mechanisms lie in several microseconds, beyond the standard actual high-performance computational limits to obtain a statistically meaningful description \cite{ali2020dynamics,zheng2017probing}, even by exploiting optimised codes for GPUs clusters \cite{casalino2021ai,lev2017string}. 
While a possible solution relies on using enhanched sampling techniques \cite{abrams2014enhanced}, the large simulation size involved in most realistic mechanisms cannot be taken easily into account. 
Thus, in the last three decades, the development of coarse-grained models has shown great scope in overcoming these limits\cite{yu2021multiscale,leong2021coarse}. The coarse-grained strategy usually aims at reducing the details of the protein structures alongside their aqueous solvent. Such a reduction shall be made with particular care to preserve the detailed description, where necessary to appropriately describe the protein structure and function\cite{monticelli2008martini,marrink2007martini,sterpone2015protein,sterpone2014opep}. 

A different approach is based on the coupling of different tools, to describe the coexistence of different time- and length- scales in  a multiscale perspective. 
Such multiscale approach could become, once assessed, a powerful biophysical computational tools, with perspective usage in drug design and biomedicine \cite{coveney2021computational,edeling2021impact,di2021three}.

The present work exploits a multiscale description based on the Lattice Boltzmann (LB) method for the solvent fluid of the aqueous combined with all-atom molecular dynamics (MD) description for the protein structures. Several multiscale approaches to coupling LB/MD have been developed so far \cite{bernaschi2019mesoscopic,mackay2013coupling,dupuis2007coupling,fyta2006multiscale,chatterji2005combining}. In this work, the coupling of the LB velocity field with MD objects is implemented via a Stokes friction term in the overlay region of the two descriptions to realise the multiscale description \cite{bernaschi2019mesoscopic}.

\begin{figure}[!ht]
\begin{center}
\includegraphics[width=0.8\linewidth]{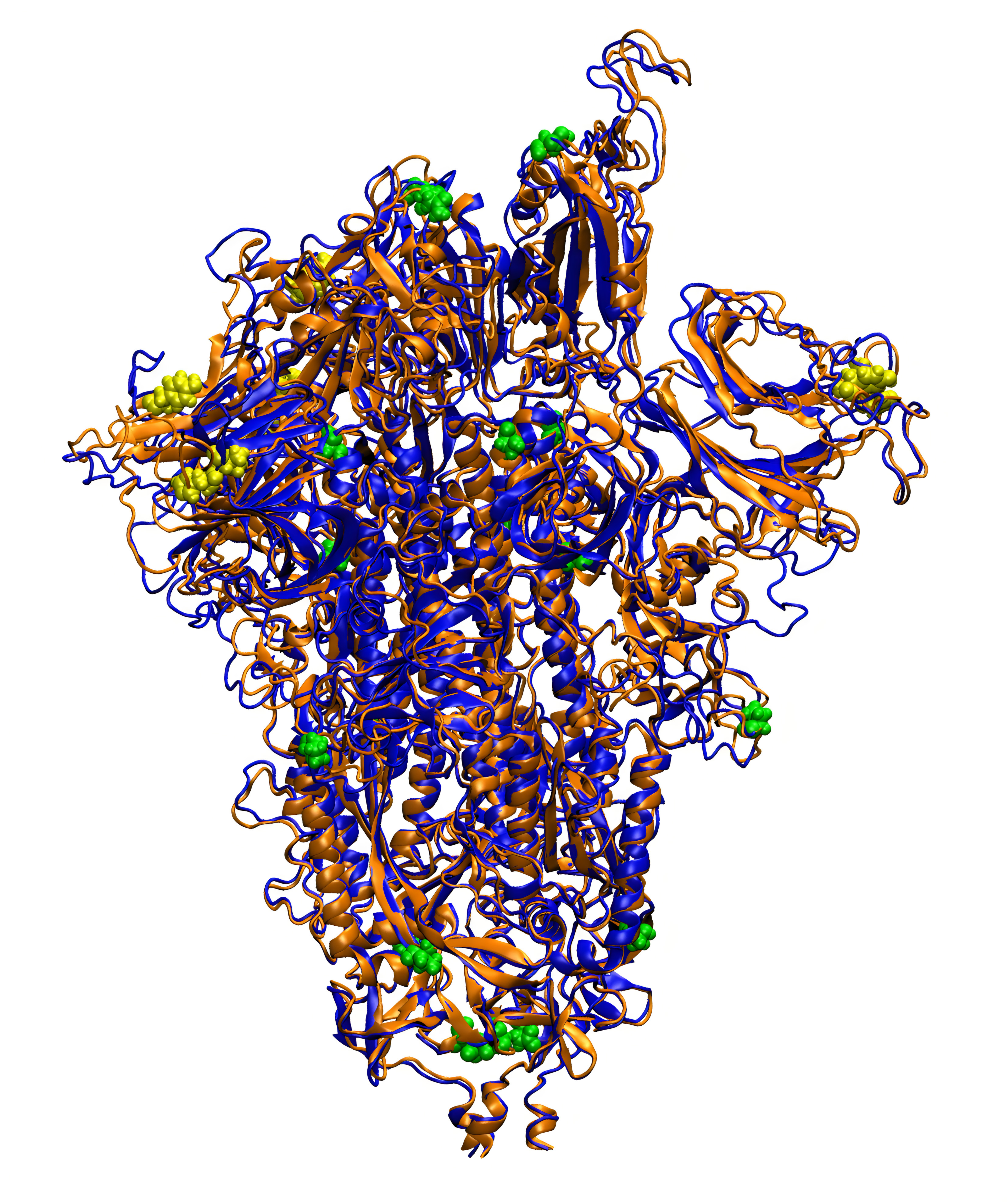}
\caption{Secondary structure of the spike protein, wild-type (blue) and Alpha variant (orange) as obtained from experimental pdb (6vsb.pdb) and homology modelling, respectively. In yellow and green, deletions and mutations of the wild-type giving the Alpha variant are highlighted. }
\label{fig:spike-prot}
\end{center}
\end{figure}

Our test case is the well known and characterized SARS-CoV-2 spike protein S, a heavily glycosylated protein anchored in the viral membrane. It is constituted by three chains, each one made of an identical primary sequence of more than 1200 amino acids of which 1146 form the extracellular domain. Each chain of the trimer is composed of two fragments: the receptor-binding fragment S1, containing the receptor-binding domain (RBD), interacting with ACE2, and the fusion fragment S2 \cite{cai2020distinct}. The S protein is cleaved by a furin-like protease at residue 686 into the S1 and S2  fragments \cite{bosch2003coronavirus}, initiating the membrane fusion process. We also study the Variant of Concern 202012/01 (lineage B.1.1.7), also commonly referred to as Alpha variant ($\alpha$-spike). 

Equilibrium properties, such as rigidity and elasticity of specific sites, are pivotal for binding and other functional activities of the viral protein \cite{cai2020distinct,spinello2020rigidity}, therefore they must be properly described within the multiscale model. We compare equilibrium state properties from LB/MD data obtained by all-atom molecular dynamics, to assess the quality and efficiency of the multiscale description. The key role of water molecules for protein structure and function is highlighted \cite{bellissent2016water}. 
In perspective, our multiscale approach could permit us to overcome the statistical sampling limitation affecting current explicit solvent atomistic description, in particular providing a unique tool to investigate large conformational changes and dynamics solvent flow effects.
However, despite main features of the protein structure and motion are reproduced in the mixed approach, a strong refinement is still necessary to properly reproduce the flexibility and fluctuations that, allowing a correct conformational sampling, are related to the proper functioning of the proteins.

\section{Methodology}
\label{sec2}
\subsection{Simulated systems}
We used cryo-EM data \cite{wrapp2020cryo,yan2020structural} to build the two models of the wild-type and mutated protein, hereafter called spike (S) and $\upalpha$-spike ($\upalpha$S), respectively. 
We started from the 6vsb.pdb structure \cite{zheng2017probing} of the spike protein in prefusion state, and we added the missing loops in the receptor-binding domain (RBD) as obtained from the 6m17.pdb structure \cite{yan2020structural}. We kept the glycans from the 6vsb.pdb structure  in these simulations. The $\upalpha$-spike has been modelled via I-Tasser \cite{yang2015tasser}, using the wild-type as template, by including in the sequence \cite{meng2021recurrent} three deletions: $\Updelta$H69/$\Updelta$V70 and $\Updelta$Y144, and six mutations;   N501Y, A570D, P681H, T716I, S982A, and D1118H (see Figure \ref{fig:spike-prot}).  
The protonation states have been calculated via Playmolecule webserver \cite{marti2017playmolecule}, based on PROPKA 3.1 \cite{olsson2011propka} to determine protein pKa values, and on PDBTOPQR 2.1 \cite{dolinsky2004pdb2pqr} to optimize the protein for favourable hydrogen bonding. 
The cell of the LB/MD systems is a cubic box of 19.2 nm side length, surrounding the all-atom glycosylated-protein (53k atoms). The same box size is used for the all-atom simulations (676k atoms). The cell is built and neutralized (100 nM solution) via the Solvate and Ionize plugins included in Visual Molecular Dynamics (VMD) \cite{humphrey1996vmd}.

\subsection{Coupled Lattice-Boltzmann and Molecular Dynamics (LB/MD)}
LB/MD simulations have been performed with LAMMPS (stable release 3 March 2020) \cite{plimpton1995fast} on a cluster based on Intel(R) Xeon(R) Platinum 8160 with 24 cores and two CPUs per computing node. 

The two systems (wild-type and mutant) endured an initial 10000 steps of conjugate gradient minimization and were afterwards equilibrated in an NVT ensemble, with the temperature increased to 310 K over 2.0 nanoseconds of MD simulations. Hence, both systems were evolved in time for 1 microsecond (NVT, 310 K). The direct summation method was exploited to assess the Coulomb interactions. In particular, the additional screening of the solvent was modelled by a Coulomb correction for implicit solvent interactions which exploits a distance-dependent dielectric permittivity, scaling with an additional 1/r term included in the Coulomb formula. The cut-off of the intermolecular interactions was set to 7 $\text{\AA}$ corresponding to the Bjerrum length in water \cite{micka1999strongly}. The CHARMM36 force field \cite{huang2013charmm36} has been used to model inter- and intra-molecular interactions, including the glycan and N-linked glycan bond descriptions. 

The aqueous solvent is described by a specific mesoscale technique, known as the Lattice Boltzmann (LB) method, namely a minimal lattice version of the Boltzmann equation. The LB approach allows modelling the dynamic behaviour of fluid flows without directly solving the Navier-Stokes equations of continuum fluid mechanics. In this framework, the solvent is treated via a fictitious ensemble of particles, whose motion and interactions are confined to a regular space-time lattice. The dramatic reduction of the degrees of freedom associated with the velocity space is the main advantage of the LB approach. Thus, the solvent is described in terms of probability to find a certain quantity of solvent particle at position $\vec{r}$ and time $t$ moving with velocity $\vec{c}_i$ along a possible grid direction. In the LB approach, the particle collisions are represented through a relaxation to the local equilibrium. Here, we rely on the simplest form of the collision operator that is the celebrated Bhatnagar-Gross-Krook operator, where the operator is a simple single-time relaxation term \cite{succi2018lattice}.

The standard LB scheme in single-relaxation time (BGK) form reads as follows:
\begin{equation} \label{LBE}
f_{i} \left(\vec{r}+\vec{c}_{i}\Delta t,\,t+\Delta t\right) =f_{i}\left(\vec{r},\,t\right)+\omega \left[ f_{i}\left(\vec{r},\,t\right) - f_{i}^{eq}\left(\vec{r},\,t\right) \right]+S_i\left(\vec{r},\,t\right),
\end{equation}
where $f_i$ is the discrete Boltzmann distribution associated with the discrete velocity $\vec{c}_i$, with $i=0, b$ running over the discrete lattice, in our case 19-speed lattices, commonly denoted D3Q19. In \ref{LBE}, the relaxation frequency $\omega$ is used to set the kinematic viscosity $\nu$ of the fluid by the relation $\omega=2/(6\nu+1)$, while $f_{i}^{eq}$ is the lattice local equilibrium, basically the local Maxwell-Boltzmann distribution truncated to the second order in the Mach number. Mass density and mass flow are obtained in terms of moments of the distribution functions: 
\begin{eqnarray} \label{moments}
\rho\left(\vec{r},\,t\right) = \sum_i f_{i}\left(\vec{r},\,t\right) \\
\rho \left(\vec{r},\,t\right)\vec{u}\left(\vec{r},\,t\right) = \sum_i f_{i}\left(\vec{r},\,t\right) \vec{c}_{i}
\end{eqnarray}
where $m$ denotes a scaling mass factor set to obtain the correct water density of 993.4 kg/m$^3$ at 1 atm and 310 K. 

It is worth to highlight that, denoted $T$ the finite temperature of the fluid, the standard LB algorithm does not include the thermal fluctuations which produce spontaneous local stresses in the fluid following the fluctuation dissipation relation as $<s_{\alpha \beta}(\vec{r},t) s_{\gamma\nu}(\vec{r}',t')>=2k_BT\eta_{\alpha\beta\gamma\nu}\delta(\vec{r}-\vec{r}')\delta(t-t')$, where $s_{\alpha \beta}$ is the fluctuating stress tensor, $\eta_{\alpha\beta\gamma\nu}$ is a tensor of viscosities and $k_B$ denotes the Boltzmann constant.
Following the seminal work by Adhikari et al. \cite{adhikari2005fluctuating}, the forcing term in Eq. \ref{LBE} is here generalised to $S_i\left(\vec{r},\,t\right)=p_i+\xi_i$, where $p_i$ counts for the external and coupling forces (see below), while $\xi_i$ represents a set of random noise terms modelling the fluctuating stress tensor compensated by the dissipation and preserving the mass and momentum conservation in the system. Further details of the actual fluctuating LB implementation are reported in references \cite{mackay2013hydrodynamic,ollila2011fluctuating}. 
Thus, as a result of the thermal noise implementation, the fluid behaves as a heat bath for the MD particle, as already observed in the literature \cite{sterpone2015protein,mackay2013coupling,dunweg2009lattice}.

The fluid and particles are coupled as follows. The effect of the particle on the surrounding fluid is modelled via a friction force term, $\vec{F}_{nj}=\gamma(\vec{\upsilon}_n-\vec{u}_j)$, where $\vec{\upsilon}_n$ denotes the particle velocity and $\vec{u}_j$ the fluid velocity at the particle position obtained by a linear interpolation over the nearest eight lattice points. The coupling force is then added to Eq. \ref{LBE} by the extra force term $S\left(\vec{r},\,t\right)$. Hence, an equal and opposite force is applied to the particle to model the counterpart of the coupling term (from the fluid to the particle). Following the literature \cite{sterpone2015protein}, the friction coefficient $\gamma$ is taken equal to  0.1 fs$^{-1}$, while the kinematic viscosity was set equal to 0.07 $\text{\AA}^2/\text{fs}$  corresponding to the water kinematic viscosity at 310 K. The LB scheme is evolved in time step by step with the MD integration scheme, with a timestep equal to 2.0 femtoseconds.

\subsubsection{All-atom molecular dynamics (AA-MD)}
AA-MD simulations have been performed with GROMACS-2020.6 \cite{abraham2015gromacs} on a cluster based on IBM Power9 architecture and Volta NVIDIA GPUs. Both spike and $\upalpha$-spike endured an initial 10000 steps of conjugate gradient minimization. Hence, the systems were equilibrated up to 310 K over 2.0 nanoseconds, as in the LB/MD simulations. The MD simulations last 1 microsecond (NPT, 310 K, 1 atm). The simulation timestep is 1.6 femtoseconds. Periodic boundary conditions were used, with particle-mesh Ewald long-range electrostatics, using a grid spacing of 1.5 $\text{\AA}$ along with a fourth-order B-spline charge interpolation scheme. Both the Coulomb and Lennard-Jones interactions use a cut-off of 12 $\text{\AA}$ with a force switching function acting from 10 $\text{\AA}$ \cite{steinbach1994new}. The CHARMM36 force field \cite{huang2013charmm36} has been used to model inter- and intra-molecular interactions, including the glycan and N-linked glycan bond descriptions. The water is described via the TIP3P water model as implemented in CHARMM \cite{mackerell1998all} which specifies a 3-site rigid water molecule with Lennard-Jones parameters and charges assigned to each of the 3 atoms.

\section{Results}
We present in this section some quantities, normally used to assess protein structures, analyzed for the two used methods and for the two considered variants.
We also compare the AA-MD wild-type results with data available in literature. We can anticipate here that our model properly reproduces main features of the protein, as obtained from similar all-atom or coarse grained calculations with explicit solvent. Then, we highlight the main structural and dynamics differences between wild-type and mutated spike, mainly in terms of rigidity, fluctuations and correlation, and relate them, when possible, to aminoacids mutations. Finally, the main objective of this work, we analyze to which extent the features, characterizing the protein, are properly reproduced by the hybrid LB-MD description, and discuss the motivation for possible differences.

\begin{figure}[!ht]
\begin{center}
\includegraphics[width=0.8\linewidth]{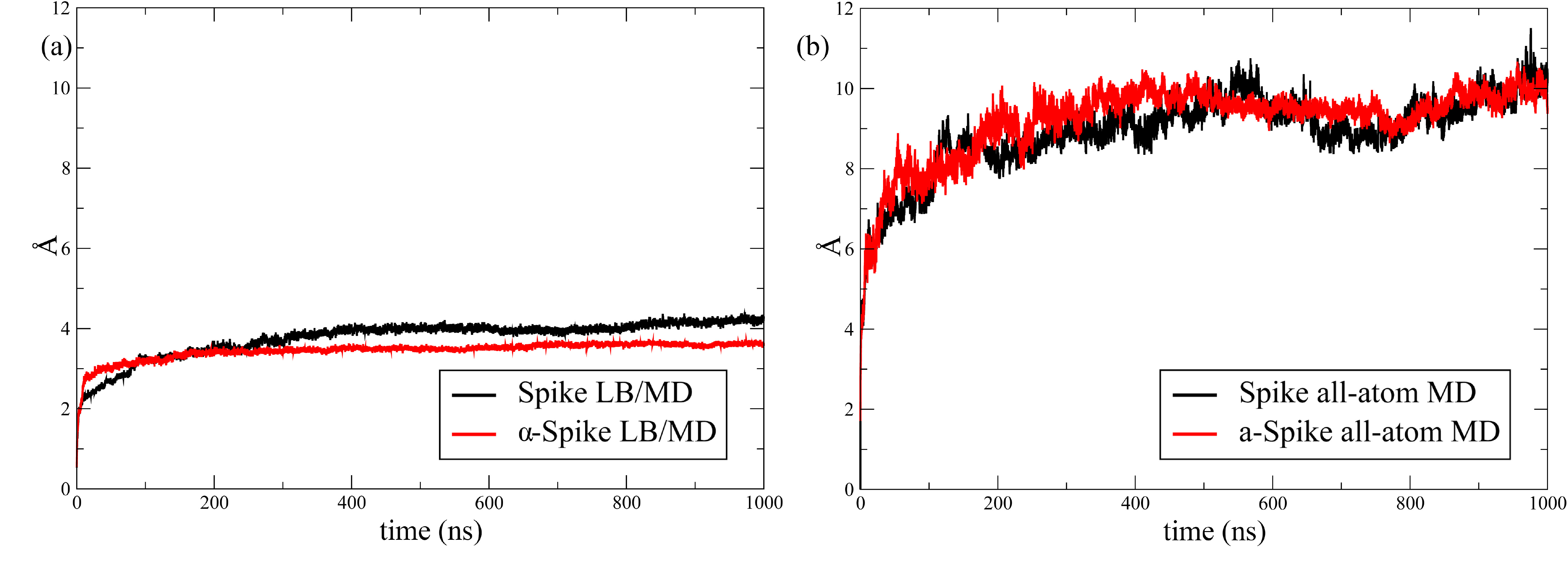}
\caption{Root Mean Square Deviation of the C$\upalpha$ atoms, for LB-MD (left panel) and AA-MD (right panel) simulations, for the 1 $\mu$s long trajectories. Slightly differences are observed between wild and mutated. The intensity decrease for the LB-MD case is related to the lower mobility of the protein in the mesoscopic solvent with respect to the explicit solvent.}
\label{fig:rmsd}
\end{center}
\end{figure}

\begin{figure}[!ht]
    \begin{center}
    \includegraphics[width=0.8\linewidth]{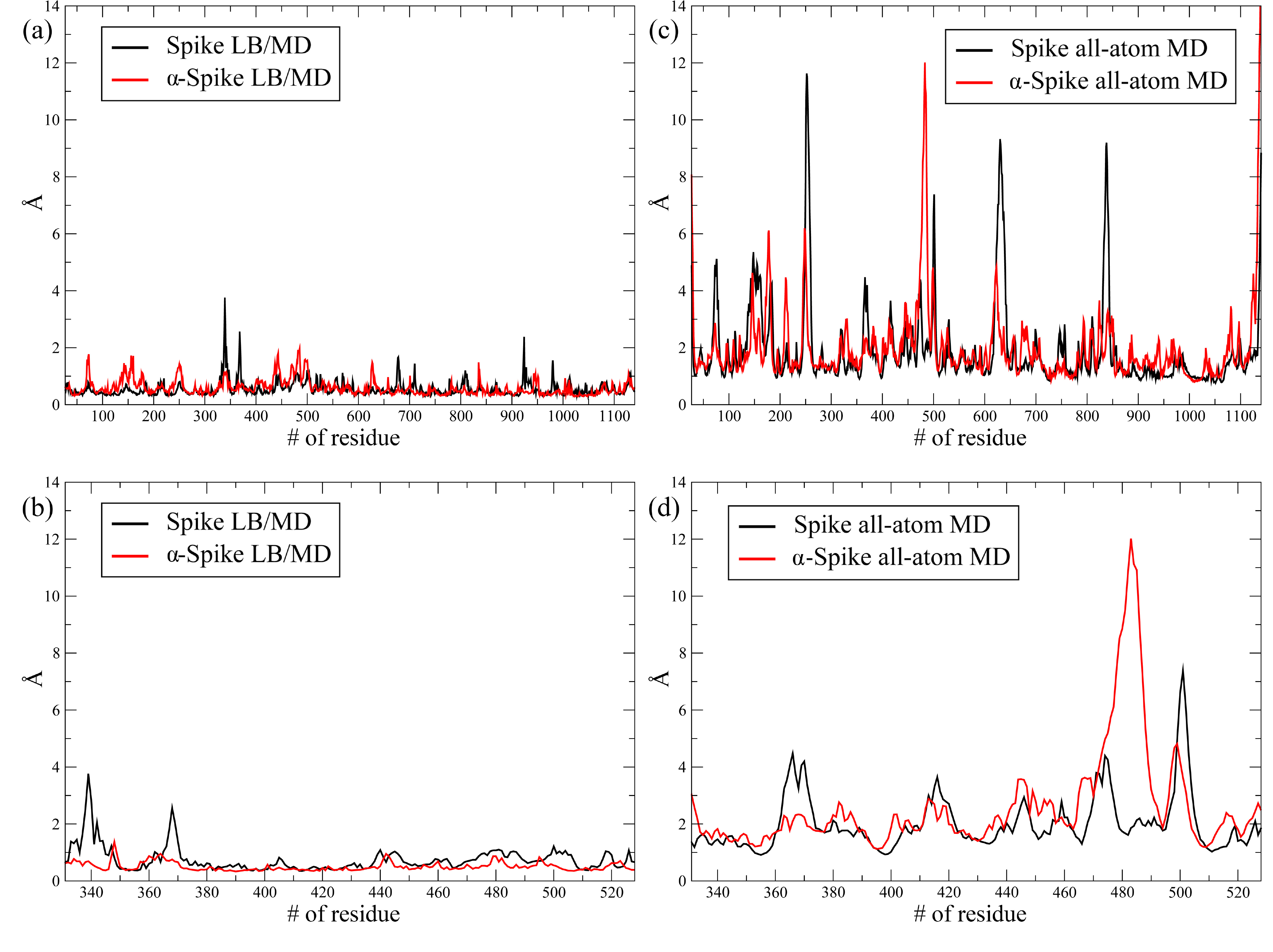}
    \end{center}
\caption{Root Mean Square Fluctuations (C$\upalpha$ atoms, RBD) for LB-MD (left panel) and AA-MD (right panel) simulations of chain A. Top panel, full chain A. Bottom panel, RBD of chain A. The initial 500 nanoseconds of the simulations were discarded. Slightly differences are observed between wild and mutated. The significant intensity decrease in the LB/MD is due to the low protein mobility. This result, expected on qualitative grounds, to the best of our knowledge was never inspected on quantitative grounds before.}
\label{fig:rmsf}
\end{figure}

The Root Mean Square Deviation (RMSD) and the Root Mean Square Fluctuations (RMSFs) of AA-MD and LB-MD for spike and $\upalpha$-spike are reported in Figs.\ref{fig:rmsd})-\ref{fig:rmsf}). 
The RMSD (see Figure \ref{fig:rmsd}) is shown for the full trajectory, lasting 1 $\mu$s.

In the AA-MD case, the RMSD of both structures are similar, with a large initial deviation from the cryo-EM structure, and a most stable behaviour from 200 ns to 1000 ns, with oscillations in the region 8\AA-10\AA. 
The RMSD analyzed per chains (not shown) has similar behaviours for chains A and B when native and mutated are compared. Chain C of the mutated relaxes faster than chain C of the native. 
Moving to the LB-MD results, the first clear information is the stability of the structure with respect to the cryo-EM reference. The overall change is 3.5\AA-4\AA, with the mutated reaching a stable plateau faster than the native structure. Chain A (not shown) of the mutated has RMSD values larger than the native, of almost 1\AA, while the opposite is true for chain B, whose native RMSD has a plateau at 5\AA, much higher than the mutated plateau at 2\AA. Chain C is similar for the two structures. 
Differences among RMSD evaluated for different chains have been reported\cite{ghorbani2021exploring}, in agreement with our findings. 
The RMSD values of the wild-type in all-atom description is similar to results obtained with a simular approach \cite{ghorbani2021exploring}, but larger that other reference data\cite{khan2021preliminary,raghuvamsi2021sars}. However, these differences could depend on the configuration used as reference in papers\cite{khan2021preliminary,raghuvamsi2021sars,ghorbani2021exploring}. In our case, given the overall size of the protein, containing 3 chains of more than 1000 residues each, and the fact that our reference structure is the model as obtained from cryo-EM and homology modeling, the RMSD value is reasonable. At variance with the mutations considered in Ref.\cite{khan2021preliminary}, which induce larger deviations with respect to the native, in the case of the $\alpha$-spike here studied the RMSD over the whole structure is comparable or lower than the native protein. 

In any case, RMSD has to be considered, more than for its absolute values, for its trend, to verify that the structure is not drifting and that equilibrium is reached. 

Overall, the conformational equilibration of the AA-MD is quite slow, with large changes from the initial models as obtained from cryo-EM data and homology modelling. The relaxation is much faster in LB-MD, and changes are smaller with respect to starting structure. Also, RMSD fluctuations are larger in AA-MD than in LB-MD. 
The observed differences are partially ascribable to the intrinsic rigidity of the LB-MD description. Also, we cannot exclude a role of the starting cryo-EM structure which could represent a micro-state of the protein conformational space far from the equilibrium. The long drift in the RMSD profile of the AA-MD simulations for spike and $\upalpha$-spike hints that the starting protein structure, as obtained from cryo-EM, is not close to the equilibrium structure of the open state. 

To use a good quality sampling, we use to evaluate the following quantities the last 500 ns of the four trajectories, disregarding the first half of each trajectory.

The secondary structures of the spike and $\upalpha$-spike proteins were analyzed by the program STRIDE \cite{frishman1995knowledge}, taking into consideration 500 configurations placed at one nanosecond of interval in the last 500 ns of the trajectories. In both the AA-MD and LB-MD systems, we do not observe a substantial difference between spike and $\upalpha$-spike. Nonetheless, we observe a slight distortion of the $\alpha$-helix in the LB-MD simulations as already observed in the literature \cite{zhou2003free,calimet2001protein}. In particular, while the secondary structure in the AA-MD simulations contains about 20\% of the residues in $\alpha$-helix, the program STRIDE classifies almost the entire $\alpha$-helix structures in the LB-MD trajectories as turn units continuously repeated in several segments of the main chain, forming distorted helices. Instead, the classification of $\beta$-sheet conformations appears to be preserved in both AA-MD and LB-MD systems, with an average value of around 20\%. The percentage of coils, denoting the degree of disorder in the secondary structure, is not increased in the LB-MD, meaning that the overall structure is preserved even in presence of the mesoscopic solvent.

The protein rigidity in the LB-MD description is confirmed by the RMSFs (see Figure \ref{fig:rmsf}, results shown for chain A, that corresponds to the RBD in open conformation). The RMSFs of the individual chains (only chain A shown here) in AA-MD native are comparable to similar data from all atom or coarse grained simulations \cite{raghuvamsi2021sars,verkhivker2020molecular}. We highlight the three peaks in the NTD region (up to 300)\cite{raghuvamsi2021sars,verkhivker2020molecular}, and the main peak in the region 476-486 of the RBD\cite{raghuvamsi2021sars,singh2021serine}, containing the S477 whose role is pivotal in the interaction with human ACE2. The cleavage site, at 682-689, is not flexible in the chain A, with RBD in open conformation, while it is mobile in the other two considered chains (data not shown). 
The most striking difference between spike and $\upalpha$-spike is in the region 476-486 of the RBD in chains A and B (latter not shown). In chain A, the mutated region is much more flexible, while the rest of the sequence is less mobile. In chain A, the 476-486 region is more flexible in the native, and in general the mutated structure is less flexible but around 850. 

\begin{figure}[!ht]
\begin{center}
\includegraphics[width=0.8\linewidth]{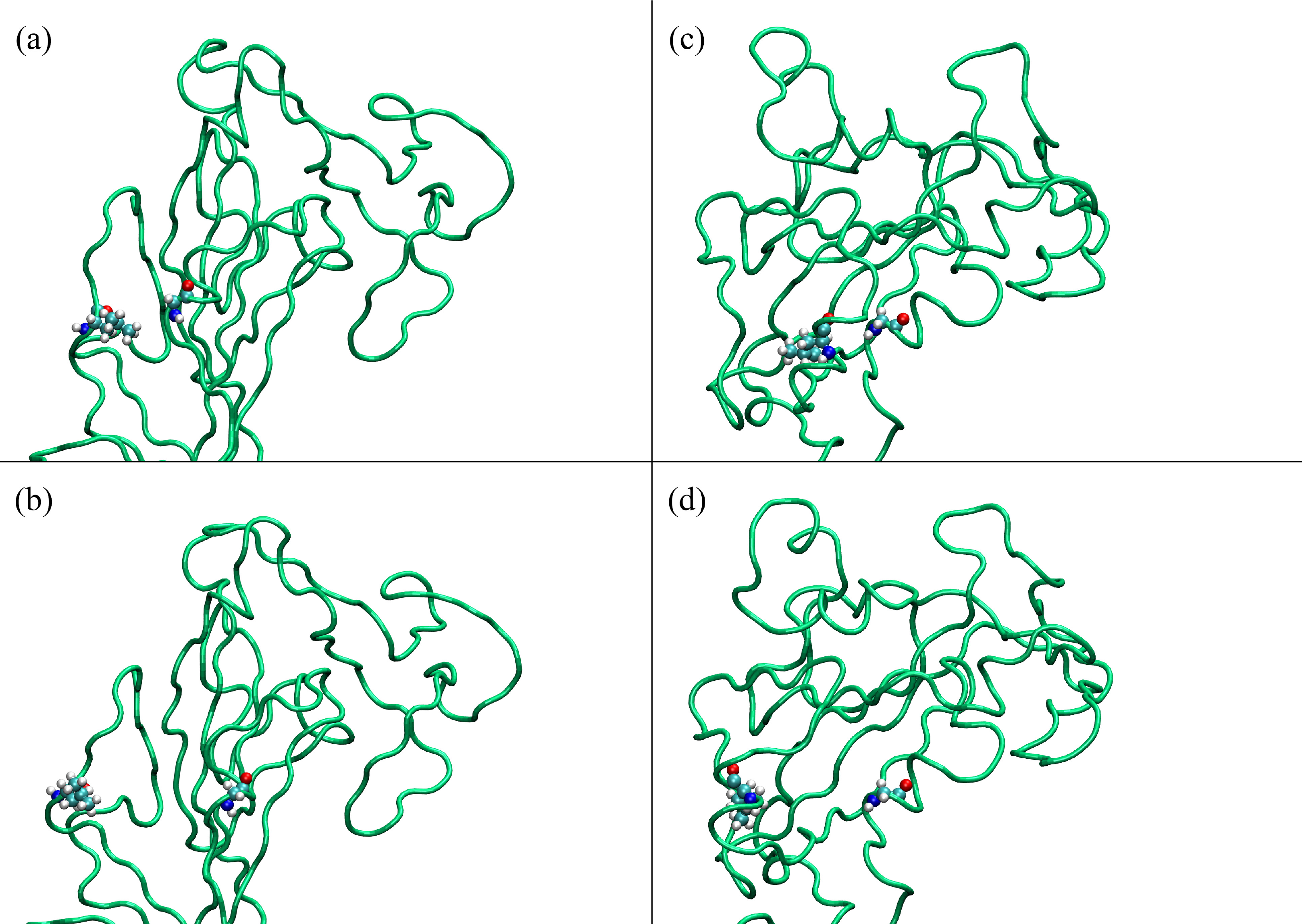}
\caption{Receptor binding domain (RBD) of the chain A in the wild-type spike protein showing the fluctuations of the residues G339 and L368 highlighted in atomistic representation for the LB/MD simulation (panels a, b) and the AA/MD simulation (panels c, d).}
\label{fig:diff-fluct}
\end{center}
\end{figure}

Fluctuations in LB-MD are less intense than in AA-MD. 
Importantly, in the RBD region, main features are reproduced by LB-MD, in particular the recognition of the loop L3 (residues from 475 to 487), of interest because related to a stable interaction with ACE2 \cite{spinello2020rigidity,singh2021serine,ngo2021identifying}, and also because of the high frequency of mutations observed in this region. 
In general, LB-MD is able to obtain a qualitative agreement with AA-MD, with fluctuations observed in the same regions. Nonetheless, there are occasional peaks in the LB-MD fluctuations over the whole sequence (at T470, Q920, S940, I980, from different chains and mainly for the native structure), comparable with the AA-MD description, that deserve to be investigated, in particular in the RBD region,    
We examine the RMSFs peaks corresponding to the residues G339 and L368, which are quite  intense in the wild-type spike protein. In Figure \ref{fig:diff-fluct}, representative snapshots of the fluctuations of the residues G339 and L368 are reported for both the LB-MD and AA-MD simulations. The residues G339 and L368 are located in two alpha-helices placed side by side, and possibly interacting. In the two simulations the amplitude of the fluctuation is  comparable, but we observe an unsound behaviour in the LB-MD case. The two residues remains close (Fig. \ref{fig:diff-fluct}(a)) for the initial 600 nanoseconds of the entire LB-MD trajectory, whereupon the two residues move further apart (Fig. \ref{fig:diff-fluct} (b)) for the remaining simulation time, probably due to the presence of an high energy barrier hindering the free fluctuation over the two positions. 
On the other hand, the residues G339 and L368 distance fluctuates in the same range much more in the AA-MD simulation (panels (c) and (d) Fig. \ref{fig:diff-fluct}). 
This behaviour emblematically demonstrates as the local free energy surface is biased by the explicit solvent, which mediates the fluctuation of the protein structure. Instead, the mesoscopic solvent promotes likely the roughness of the free-energy landscape, which restrains the free fluctuations of the structure, and promotes the presence of hindered fluctuations (bi-modal distributions) in the residue positions, as here observed in the case of the G339 and L368 residues in the LB-MD simulation. 

The effects on free energy sampling related to different models of solvent, implicit or explicit, have been widely investigated, for investigating folding processes and for studying the stability of native structures \cite{chen2021machine,kulke2018replica,shao2018assessing,zhou2003free,suenaga2003replica}. Further, it is worth highlighting that hybrid solvent approaches have also been proposed in several references \cite{onufriev2018water,roe2007secondary,okur2006improved,deng2004hydration}. In particular, these approaches exploit an explicit atomistic representation of the water molecules in the first solvation shell while the bulk solvent is modeled at the continuum level. Nonetheless, the LAMMPS code used in the present work does not yet include hybrid solvent representations.

To gain some insights into the quaternary motion of the trimer, we performed the principal component analysis (PCA) of the internal motion involving the linked S1 and S2 fragments and also evaluated the cross-correlation matrix (CCM) between pairs of C$\upalpha$. 

\begin{figure}[!ht]
\begin{center}
\includegraphics[width=0.8\linewidth]{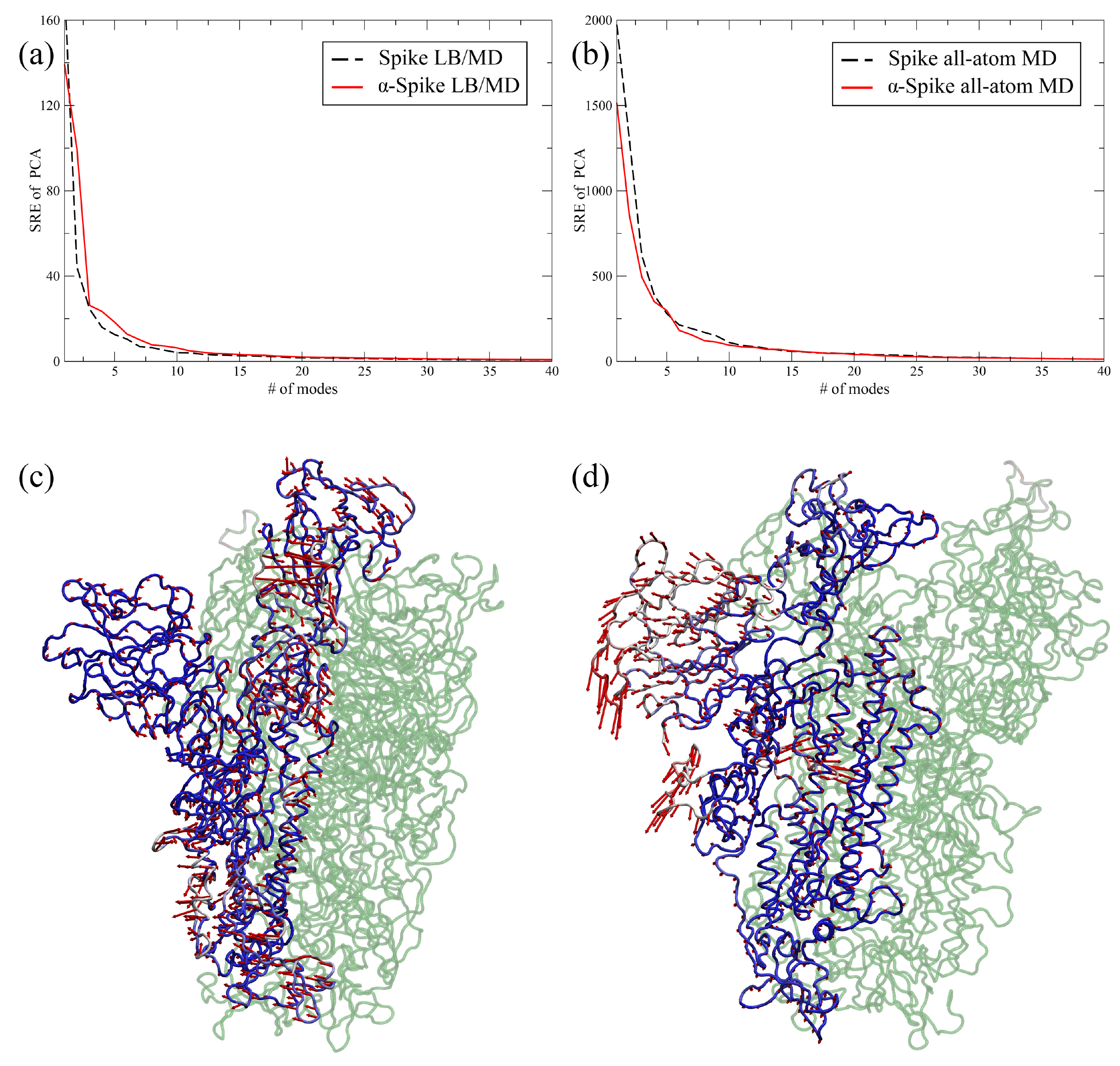}
\caption{Top panel: Square root of the eigenvalues (SRE) of the largest 40 modes from the PCA, for chain A of LB/MD (left panel) and AA/MD (right panel) simulations. Slightly differences are observed between wild and mutated. A strong intensity decrease is observed for the LB-MD cases, corresponding to the reduced mobility of the protein. Bottom panel: The main variation mode in principal component analysis (PCA) for chain A of the wild-type Spike protein in the LB/MD simulation (a) and in the AA/MD simulation (b).}
\label{fig:sre-pca}
\end{center}
\end{figure}

The PCA (see Figure \ref{fig:sre-pca}) is  quite indicative, for the sake of methods' comparison.  Focusing on Spike/$\upalpha$-Spike comparison, their description is quite similar in the framework of the same method. Obviously, due to the lower flexibility of the LB-MD protein, the size of the modes is almost 3-4 times larger in AA-MD. Moreover, also the weight of the various modes is different, meaning that different regions of the protein show different flexibility changes when the solvent is described all-atom or at the mesoscale. For example, the first mode involves mainly the NTD in the AA-MD description, and the RBD in the LB-MD model. 

The PCA analysis has been used, in a all-atom simulation, to demonstrate that the active form of the CoV-2 spike protein is more stable than the active CoV-1 spike protein \cite{kumar2021differential}, with the first normal mode associated to the motion of RBD, as here correctly described by LB. A concerted motion of NTD and RBD is observed as second normal mode. In a different simulation \cite{ghorbani2021exploring}, similar involvement of NTD and RBD in the first normal mode has been observed. 

\begin{figure}[!ht]
\begin{center}
\includegraphics[width=0.8\linewidth]{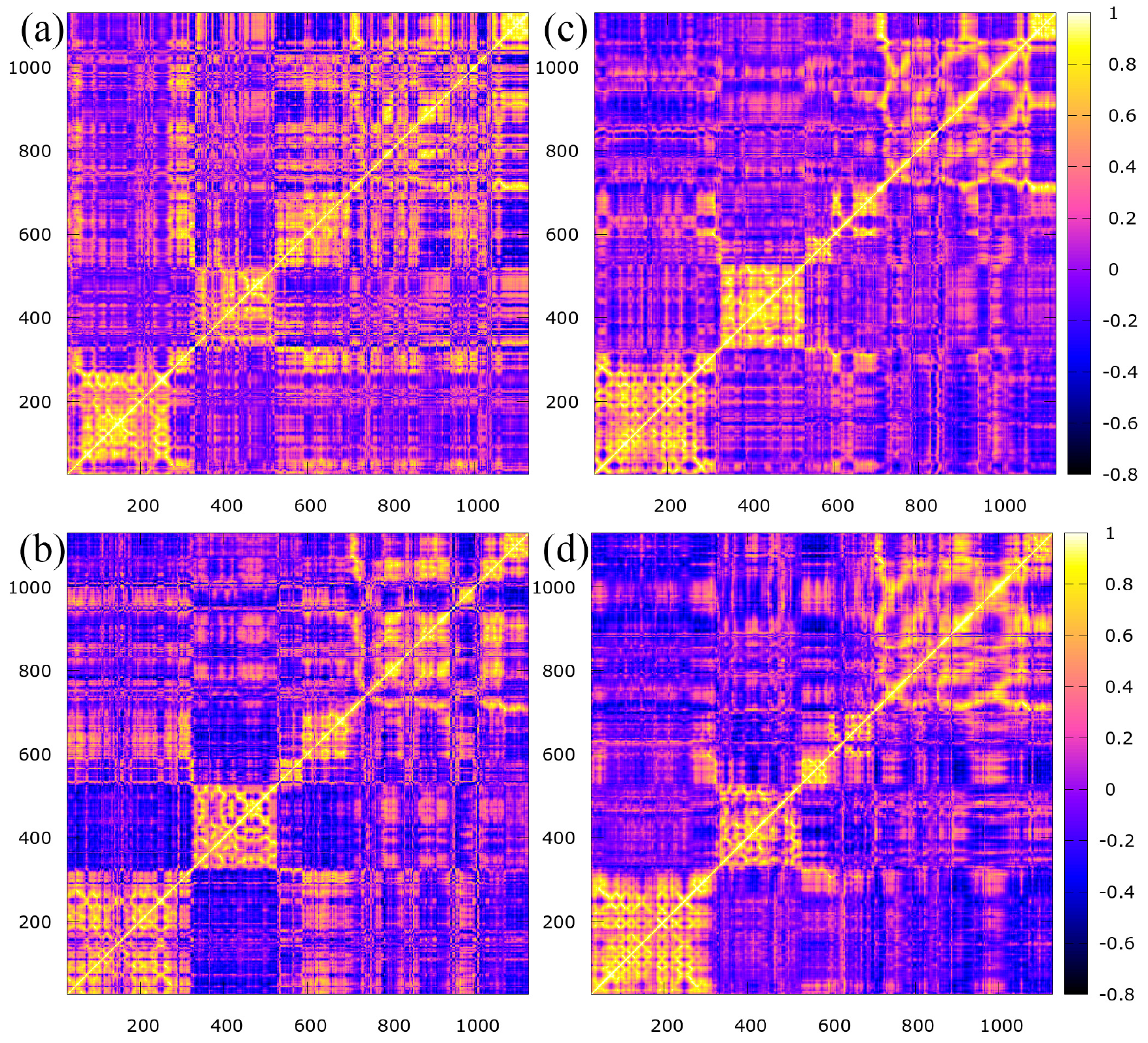}
\caption{Cross-correlation map of the cross-correlation matrix for chain A of Spike and $\upalpha$-Spike in LB/MD simulations (panels a, b), and in AA/MD simulations (panels c, d). All cases show a significant correlation block associated with residues from the RBD (residues 331-528). The higher mobility of the protein in the all-atom simulations is reflected by the richer structure of the CCM (panels c,d), yet the qualitative structure of the patterns is preserved in LB/MD. }
\label{fig:pca-cr}
\end{center}
\end{figure}

Complementary information can be obtained from the correlation (CC) maps (Figure \ref{fig:pca-cr}). At a first glance, the maps for the two methods and the two protein variants show similar macroscopic behaviour, also comparable with data in literature \cite{kumar2021differential}. 
Correlations between residues are reasonably described by LB-MD with respect to AA-MD. This result is due to the fact that intramolecular local interactions, not mediated by solvent, are kept and well described in the LB-MD method. To note that, in both methods, the $\upalpha$-Spike presents higher correlation values with respect to the wild-type, denoting that mutations indeed affect the protein structure and interaction propensity, and that these fundamental features are kept in the mesoscale description. High correlation regions are well related to protein domains.

\section{Discussion and Conclusions}

In this work, we demonstrate how some fundamental protein structural and dynamical features are correctly described in the scheme LB-MD, that couples the atomistic description of the protein to the mesoscale model for the solvent. 
However, some critical points can be clearly enucleated and quantified. 

First of all, due to the implicit solvent model, the H-bonds, both intra-protein and between protein and solvent molecules, are completely missing in the LB-MD scheme. 
The H-bonds and the interaction with water molecules are fundamental for protein stability and conformational flexibility, and also for protein-protein interactions. In this specific case, for example, the H-bonds formation is at the basis of the spike-ACE2 interaction \cite{piplani2021silico}.
Also, as a consequence of the mesoscale solvent model, the protein flexibility and conformational fluctuations are strongly inhibited. 
Conformational fluctuations are fundamental because they are related to the possibility of a protein to change structural conformation and therefore to function. For the specific case of the spike protein, the fluctuations of the L3 loop, containing S477, in the RBD, are fundamental to guarantee the spike ancoring to ACE \cite{singh2021serine}.
A reliable structural description and analysis of the RBD in the prefusion, unbound spike, is therefore a very critical step, for example to design inhibitors or antibodies to block hACE2 and RBD interactions \cite{singh2021serine}.
The motion of RBD is only partially guaranteed by the LB-MD, which provides a description intrinsically more rigid than the AA-MD.
However, LB-MD properly identifies the region as more flexible, and cross-correlated, with respect to the rest of the sequence, as shown for instance by the CC map and the RMSFs. The RMSFs, despite being quantitatively different between the two methods, show similar qualitative behaviour when comparing Spike and $\upalpha$-Spike. In particular, we observe an increased rigidity of the RBD in the $\upalpha$-Spike compared to the wild-type Spike.  The main correlations inside the sub-domains are preserved, and the large correlation of the RBD block in the CC map for the $\upalpha$-Spike reinforces the information on the stability of the mutated structure. Indeed, the capability of the $\upalpha$-Spike RBD to maintain a rigid structure can be related, from the statistical point of view, with its remarkable human-to human transmissibility \cite{spinello2020rigidity}, since the smaller fluctuations are limiting the sampling of possible RBD configuration basins to the subregion where the structure more efficiently binds to ACE2. This behaviour is somehow similar to what is observed for RBD rigidity of SARS-CoV-2 with respect to the Sars virus \cite{spinello2020rigidity}. Moreover, the same correlation trends over Spike and $\upalpha$-Spike structures are found in the LB/MD simulations, showing the capability of the multiscale approach to preserve essential information even if at a lower description level of the solvent.

Furthermore, we performed preliminary simulations (not shown) where the ACE2-RBD interaction is successfully realized also in the LB-MD approach, overcoming both the scarce loop flexibility and the lack of H-bonds. 

Overall, the comparison with the other reductionist method, the coarse grained description \cite{leong2021coarse} shows that similar results can be obtained with LB-MD. 
We investigated here only the open state of spike. Most recent investigations \cite{verkhivker2020molecular} on different conformational states show that locked and closed states are more rigid than the open state. So, the LB-MD model would be probably more effective in describing those functional states, than the more flexible open state. 
Despite its limitations, LB-MD can keep trace of structural changes in flexibility related to mutations, therefore potentially preserving the biological information related to the virus behaviour for different lineages.

Overall, from our investigation, two main results emerge: a) the {\it in nuce} potential of the multiscale approach to describe large systems, with the possibility of including flow motion effects in the protein dynamics; b) the needing for a more refined coupling scheme at the protein/solvent interface, possibly a scheme where the solvent is explicitly described only close to the protein interface, providing a reliable description of the interaction among water molecules and amino acids, a pivotal interaction in determining both equilibrium and dynamic properties of biological molecules. 
This needing still poses problems, as testified by the continuous publication of research papers on the topic \cite{chen2021machine,tao2020using,gao2020short,chakravorty2018reproducing,kulke2018replica}. Further, the LB method could also be extended to include hydrogen bond interactions directly on the lattice representation. For instance, several water-like potentials have been developed in the last decades to describe water molecules' polarity on a lattice representation. In these potentials, each cubic box could contain about one water molecule \cite{stokely2010effect,franzese2002liquid} or also a cluster of several water molecules \cite{ben2008one,lovett1969one}. Hence, an effective Hamiltonian is usually defined to model the hydrogen bond interaction in the bulk water and in the protein hydration shell \cite{bianco2017stability,bianco2017role}. As a practical example, the three-dimensional Ben-Naim \cite{ben1971statistical} water potential was included in the LB method to model hydrogen bonds in water bulk \cite{succi2014lattice}.
However, the use of coarse-grained potentials natively developed to treat the solvent implicitly makes unnecessary the use of hydrogen bond extensions in the LB model. Notable examples of implicit-solvent coarse-grained models are OPEP \cite{barroso2019opep6} and dry-Martini \cite{arnarez2015dry} force fields. In particular, the OPEP force field was successfully combined to LB method to investigate the protein unfolding under high shear rate \cite{sterpone2018molecular}. Moreover, coarse-grained force fields could allow the simulation of large system sizes, where laminar flows could be fully developed 
beyond the effects of the local thermal fluctuations due to the simultaneous presence of large structures, such as planar membranes, receptors, and viral capsids. 

To conclude, on one side our simulations demonstrate and quantitatively estimate the pivotal role of the explicit water molecules treatment to obtain a statistically reliable characterization of biological molecules. In this respect the use of the LB solvent, while computationally advantageous, does not deliver quantitatively accurate information. On the other hand, our study highlights that many structural features, important in biological activities, are preserved in the LB-MD mesoscale solvent description.

Possible future development could include a new class of LB models, capable of supporting larger fluctuations than presently possible. A promising direction along this line is the resort to higher-order sets of discrete velocities as well as extension to account properly the Hydrogen bond formations.  

\section*{Acknowledgments}
M. L., F. B., M. D., A. M., A. T. and S. S. acknowledge funding from the European Research Council under the European Union's Horizon 2020 Framework Programme (No. FP/2014-2020) ERC Grant Agreement No.739964 (COPMAT). L. C., A. L. and S. F. acknowledge the support of the International Center for Relativistic Astrophysics Network (ICRANet). M. L., L. C., A. L. and S. F. acknowledge the support of the Italian National Group for Mathematical Physics (GNFM-INdAM). All the Authors gratefully acknowledge ENEA for the availability of high-performance computing resources and support on the HPC CRESCO facility used in the LB/MD simulations, under the initiative Associazione Big Data COVID-19 Fast Track. We acknowledge CINECA Project ABRISP granted under the ISCRA initiative, for the availability of high-performance computing resources and support used in the all-atom MD simulations.

\newpage

\begin{thebibliography}{10}

\bibitem{rochaix2019dynamic}
J.-D. Rochaix, ``Dynamic modeling of a 100-million-atom organelle at the source
  of life,'' {\em Cell}, vol.~179, no.~5, pp.~1012--1014, 2019.

\bibitem{song2018mechanism}
X.~Song, M.~{\O}. Jensen, V.~Jogini, R.~A. Stein, C.-H. Lee, H.~S. Mchaourab,
  D.~E. Shaw, and E.~Gouaux, ``Mechanism of nmda receptor channel block by
  mk-801 and memantine,'' {\em Nature}, vol.~556, no.~7702, pp.~515--519, 2018.

\bibitem{gadioli2021exscalate}
D.~Gadioli, E.~Vitali, F.~Ficarelli, C.~Latini, C.~Manelfi, C.~Talarico,
  C.~Silvano, C.~Cavazzoni, G.~Palermo, and A.~R. Beccari, ``Exscalate: An
  extreme-scale in-silico virtual screening platform to evaluate 1 trillion
  compounds in 60 hours on 81 pflops supercomputers,'' {\em arXiv preprint
  arXiv:2110.11644}, 2021.

\bibitem{taka2021critical}
E.~Taka, S.~Z. Yilmaz, M.~Golcuk, C.~Kilinc, U.~Aktas, A.~Yildiz, and M.~Gur,
  ``Critical interactions between the sars-cov-2 spike glycoprotein and the
  human ace2 receptor,'' {\em The Journal of Physical Chemistry B}, 2021.

\bibitem{he2020molecular}
J.~He, H.~Tao, Y.~Yan, S.-Y. Huang, and Y.~Xiao, ``Molecular mechanism of
  evolution and human infection with sars-cov-2,'' {\em Viruses}, vol.~12,
  no.~4, p.~428, 2020.

\bibitem{armijos2020sars}
V.~Armijos-Jaramillo, J.~Yeager, C.~Muslin, and Y.~Perez-Castillo,
  ``Sars-cov-2, an evolutionary perspective of interaction with human ace2
  reveals undiscovered amino acids necessary for complex stability,'' {\em
  Evolutionary Applications}, vol.~13, no.~9, pp.~2168--2178, 2020.

\bibitem{zou2020computational}
J.~Zou, J.~Yin, L.~Fang, M.~Yang, T.~Wang, W.~Wu, M.~A. Bellucci, and P.~Zhang,
  ``Computational prediction of mutational effects on sars-cov-2 binding by
  relative free energy calculations,'' {\em Journal of chemical information and
  modeling}, vol.~60, no.~12, pp.~5794--5802, 2020.

\bibitem{brielle2020sars}
E.~S. Brielle, D.~Schneidman-Duhovny, and M.~Linial, ``The sars-cov-2 exerts a
  distinctive strategy for interacting with the ace2 human receptor,'' {\em
  Viruses}, vol.~12, no.~5, p.~497, 2020.

\bibitem{gao2021methodology}
K.~Gao, R.~Wang, J.~Chen, L.~Cheng, J.~Frishcosy, Y.~Huzumi, Y.~Qiu,
  T.~Schluckbier, and G.-W. Wei, ``Methodology-centered review of molecular
  modeling, simulation, and prediction of sars-cov-2,'' {\em arXiv preprint
  arXiv:2102.00971}, 2021.

\bibitem{muratov2021critical}
E.~N. Muratov, R.~Amaro, C.~H. Andrade, N.~Brown, S.~Ekins, D.~Fourches,
  O.~Isayev, D.~Kozakov, J.~L. Medina-Franco, K.~M. Merz, {\em et~al.}, ``A
  critical overview of computational approaches employed for covid-19 drug
  discovery,'' {\em Chemical Society Reviews}, 2021.

\bibitem{ali2020dynamics}
A.~Ali and R.~Vijayan, ``Dynamics of the ace2--sars-cov-2/sars-cov spike
  protein interface reveal unique mechanisms,'' {\em Scientific reports},
  vol.~10, no.~1, pp.~1--12, 2020.

\bibitem{zheng2017probing}
W.~Zheng, H.~Wen, G.~J. Iacobucci, and G.~K. Popescu, ``Probing the structural
  dynamics of the nmda receptor activation by coarse-grained modeling,'' {\em
  Biophysical journal}, vol.~112, no.~12, pp.~2589--2601, 2017.

\bibitem{casalino2021ai}
L.~Casalino, A.~C. Dommer, Z.~Gaieb, E.~P. Barros, T.~Sztain, S.-H. Ahn,
  A.~Trifan, A.~Brace, A.~T. Bogetti, A.~Clyde, {\em et~al.}, ``Ai-driven
  multiscale simulations illuminate mechanisms of sars-cov-2 spike dynamics,''
  {\em The International Journal of High Performance Computing Applications},
  p.~10943420211006452, 2021.

\bibitem{lev2017string}
B.~Lev, S.~Murail, F.~Poitevin, B.~A. Cromer, M.~Baaden, M.~Delarue, and T.~W.
  Allen, ``String method solution of the gating pathways for a pentameric
  ligand-gated ion channel,'' {\em Proceedings of the National Academy of
  Sciences}, vol.~114, no.~21, pp.~E4158--E4167, 2017.

\bibitem{abrams2014enhanced}
C.~Abrams and G.~Bussi, ``Enhanced sampling in molecular dynamics using
  metadynamics, replica-exchange, and temperature-acceleration,'' {\em
  Entropy}, vol.~16, no.~1, pp.~163--199, 2014.

\bibitem{yu2021multiscale}
A.~Yu, A.~J. Pak, P.~He, V.~Monje-Galvan, L.~Casalino, Z.~Gaieb, A.~C. Dommer,
  R.~E. Amaro, and G.~A. Voth, ``A multiscale coarse-grained model of the
  sars-cov-2 virion,'' {\em Biophysical journal}, vol.~120, no.~6,
  pp.~1097--1104, 2021.

\bibitem{leong2021coarse}
T.~Leong, C.~Voleti, and Z.~Peng, ``Coarse-grained modeling of coronavirus
  spike proteins and ace2 receptors,'' {\em Frontiers in Physics}, vol.~9,
  2021.

\bibitem{monticelli2008martini}
L.~Monticelli, S.~K. Kandasamy, X.~Periole, R.~G. Larson, D.~P. Tieleman, and
  S.-J. Marrink, ``The martini coarse-grained force field: extension to
  proteins,'' {\em Journal of chemical theory and computation}, vol.~4, no.~5,
  pp.~819--834, 2008.

\bibitem{marrink2007martini}
S.~J. Marrink, H.~J. Risselada, S.~Yefimov, D.~P. Tieleman, and A.~H. De~Vries,
  ``The martini force field: coarse grained model for biomolecular
  simulations,'' {\em The journal of physical chemistry B}, vol.~111, no.~27,
  pp.~7812--7824, 2007.

\bibitem{sterpone2015protein}
F.~Sterpone, P.~Derreumaux, and S.~Melchionna, ``Protein simulations in fluids:
  Coupling the opep coarse-grained force field with hydrodynamics,'' {\em
  Journal of chemical theory and computation}, vol.~11, no.~4, pp.~1843--1853,
  2015.

\bibitem{sterpone2014opep}
F.~Sterpone, S.~Melchionna, P.~Tuffery, S.~Pasquali, N.~Mousseau,
  T.~Cragnolini, Y.~Chebaro, J.-F. St-Pierre, M.~Kalimeri, A.~Barducci, {\em
  et~al.}, ``The opep protein model: from single molecules, amyloid formation,
  crowding and hydrodynamics to dna/rna systems,'' {\em Chemical Society
  reviews}, vol.~43, no.~13, pp.~4871--4893, 2014.

\bibitem{coveney2021computational}
P.~V. Coveney, A.~Hoekstra, B.~Rodriguez, and M.~Viceconti, ``Computational
  biomedicine. part ii: organs and systems,'' {\em Interface Focus}, vol.~11,
  no.~1, p.~20200082, 2021.

\bibitem{edeling2021impact}
W.~Edeling, H.~Arabnejad, R.~Sinclair, D.~Suleimenova, K.~Gopalakrishnan,
  B.~Bosak, D.~Groen, I.~Mahmood, D.~Crommelin, and P.~V. Coveney, ``The impact
  of uncertainty on predictions of the covidsim epidemiological code,'' {\em
  Nature Computational Science}, vol.~1, no.~2, pp.~128--135, 2021.

\bibitem{di2021three}
F.~Di~Palma, S.~Succi, F.~Sterpone, M.~Lauricella, F.~P{\'e}rot, and
  S.~Melchionna, ``Three-stage multiscale modelling of the nmda
  neuroreceptor,'' {\em Molecular Physics}, vol.~119, no.~19-20, p.~e1928312,
  2021.

\bibitem{bernaschi2019mesoscopic}
M.~Bernaschi, S.~Melchionna, and S.~Succi, ``Mesoscopic simulations at the
  physics-chemistry-biology interface,'' {\em Reviews of Modern Physics},
  vol.~91, no.~2, p.~025004, 2019.

\bibitem{mackay2013coupling}
F.~Mackay and C.~Denniston, ``Coupling md particles to a lattice-boltzmann
  fluid through the use of conservative forces,'' {\em Journal of Computational
  Physics}, vol.~237, pp.~289--298, 2013.

\bibitem{dupuis2007coupling}
A.~Dupuis, E.~Kotsalis, and P.~Koumoutsakos, ``Coupling lattice boltzmann and
  molecular dynamics models for dense fluids,'' {\em Physical Review E},
  vol.~75, no.~4, p.~046704, 2007.

\bibitem{fyta2006multiscale}
M.~G. Fyta, S.~Melchionna, E.~Kaxiras, and S.~Succi, ``Multiscale coupling of
  molecular dynamics and hydrodynamics: application to dna translocation
  through a nanopore,'' {\em Multiscale Modeling \& Simulation}, vol.~5, no.~4,
  pp.~1156--1173, 2006.

\bibitem{chatterji2005combining}
A.~Chatterji and J.~Horbach, ``Combining molecular dynamics with lattice
  boltzmann: A hybrid method for the simulation of (charged) colloidal
  systems,'' {\em The Journal of chemical physics}, vol.~122, no.~18,
  p.~184903, 2005.

\bibitem{cai2020distinct}
Y.~Cai, J.~Zhang, T.~Xiao, H.~Peng, S.~M. Sterling, R.~M. Walsh, S.~Rawson,
  S.~Rits-Volloch, and B.~Chen, ``Distinct conformational states of sars-cov-2
  spike protein,'' {\em Science}, vol.~369, no.~6511, pp.~1586--1592, 2020.

\bibitem{bosch2003coronavirus}
B.~J. Bosch, R.~Van~der Zee, C.~A. De~Haan, and P.~J. Rottier, ``The
  coronavirus spike protein is a class i virus fusion protein: structural and
  functional characterization of the fusion core complex,'' {\em Journal of
  virology}, vol.~77, no.~16, pp.~8801--8811, 2003.

\bibitem{spinello2020rigidity}
A.~Spinello, A.~Saltalamacchia, and A.~Magistrato, ``Is the rigidity of
  sars-cov-2 spike receptor-binding motif the hallmark for its enhanced
  infectivity? insights from all-atom simulations,'' {\em The journal of
  physical chemistry letters}, vol.~11, no.~12, pp.~4785--4790, 2020.

\bibitem{bellissent2016water}
M.-C. Bellissent-Funel, A.~Hassanali, M.~Havenith, R.~Henchman, P.~Pohl,
  F.~Sterpone, D.~Van Der~Spoel, Y.~Xu, and A.~E. Garcia, ``Water determines
  the structure and dynamics of proteins,'' {\em Chemical reviews}, vol.~116,
  no.~13, pp.~7673--7697, 2016.

\bibitem{wrapp2020cryo}
D.~Wrapp, N.~Wang, K.~S. Corbett, J.~A. Goldsmith, C.-L. Hsieh, O.~Abiona,
  B.~S. Graham, and J.~S. McLellan, ``Cryo-em structure of the 2019-ncov spike
  in the prefusion conformation,'' {\em Science}, vol.~367, no.~6483,
  pp.~1260--1263, 2020.

\bibitem{yan2020structural}
R.~Yan, Y.~Zhang, Y.~Li, L.~Xia, Y.~Guo, and Q.~Zhou, ``Structural basis for
  the recognition of sars-cov-2 by full-length human ace2,'' {\em Science},
  vol.~367, no.~6485, pp.~1444--1448, 2020.

\bibitem{yang2015tasser}
J.~Yang, R.~Yan, A.~Roy, D.~Xu, J.~Poisson, and Y.~Zhang, ``The i-tasser suite:
  protein structure and function prediction,'' {\em Nature methods}, vol.~12,
  no.~1, pp.~7--8, 2015.

\bibitem{meng2021recurrent}
B.~Meng, S.~A. Kemp, G.~Papa, R.~Datir, I.~A. Ferreira, S.~Marelli, W.~T.
  Harvey, S.~Lytras, A.~Mohamed, G.~Gallo, {\em et~al.}, ``Recurrent emergence
  of sars-cov-2 spike deletion h69/v70 and its role in the alpha variant b.
  1.1. 7,'' {\em Cell reports}, vol.~35, no.~13, p.~109292, 2021.

\bibitem{marti2017playmolecule}
G.~Mart\'{i}nez-Rosell, T.~Giorgino, and G.~De~Fabritiis, ``Playmolecule
  proteinprepare: A web application for protein preparation for molecular
  dynamics simulations,'' {\em Journal of Chemical Information and Modeling},
  vol.~57, no.~7, pp.~1511--1516, 2017.

\bibitem{olsson2011propka}
M.~H.~M. Olsson, C.~R. S{\o}ndergaard, M.~Rostkowski, and J.~H. Jensen,
  ``Propka3: Consistent treatment of internal and surface residues in empirical
  pka predictions,'' {\em Journal of Chemical Information and Modeling},
  vol.~7, no.~2, pp.~525--537, 2011.

\bibitem{dolinsky2004pdb2pqr}
T.~J. Dolinsky, J.~E. Nielsen, J.~A. McCammon, and N.~A. Baker, ``Pdb2pqr: an
  automated pipeline for the setup of poisson--boltzmann electrostatics
  calculations,'' {\em Nucleic acids research}, vol.~32, no.~suppl\_2,
  pp.~W665--W667, 2004.

\bibitem{humphrey1996vmd}
W.~Humphrey, A.~Dalke, and K.~Schulten, ``Vmd: visual molecular dynamics,''
  {\em Journal of molecular graphics}, vol.~14, no.~1, pp.~33--38, 1996.

\bibitem{plimpton1995fast}
S.~Plimpton, ``Fast parallel algorithms for short-range molecular dynamics,''
  {\em Journal of computational physics}, vol.~117, no.~1, pp.~1--19, 1995.

\bibitem{micka1999strongly}
U.~Micka, C.~Holm, and K.~Kremer, ``Strongly charged, flexible polyelectrolytes
  in poor solvents: molecular dynamics simulations,'' {\em Langmuir}, vol.~15,
  no.~12, pp.~4033--4044, 1999.

\bibitem{huang2013charmm36}
J.~Huang and A.~D. MacKerell~Jr, ``Charmm36 all-atom additive protein force
  field: Validation based on comparison to nmr data,'' {\em Journal of
  computational chemistry}, vol.~34, no.~25, pp.~2135--2145, 2013.

\bibitem{succi2018lattice}
S.~Succi and S.~Succi, {\em The lattice Boltzmann equation: for complex states
  of flowing matter}.
\newblock Oxford University Press, 2018.

\bibitem{adhikari2005fluctuating}
R.~Adhikari, K.~Stratford, M.~Cates, and A.~Wagner, ``Fluctuating lattice
  boltzmann,'' {\em EPL (Europhysics Letters)}, vol.~71, no.~3, p.~473, 2005.

\bibitem{mackay2013hydrodynamic}
F.~Mackay, S.~T. Ollila, and C.~Denniston, ``Hydrodynamic forces implemented
  into lammps through a lattice-boltzmann fluid,'' {\em Computer Physics
  Communications}, vol.~184, no.~8, pp.~2021--2031, 2013.

\bibitem{ollila2011fluctuating}
S.~T. Ollila, C.~Denniston, M.~Karttunen, and T.~Ala-Nissila, ``Fluctuating
  lattice-boltzmann model for complex fluids,'' {\em The Journal of chemical
  physics}, vol.~134, no.~6, p.~064902, 2011.

\bibitem{dunweg2009lattice}
B.~D{\"u}nweg and A.~J. Ladd, ``Lattice boltzmann simulations of soft matter
  systems,'' {\em Advanced computer simulation approaches for soft matter
  sciences III}, pp.~89--166, 2009.

\bibitem{abraham2015gromacs}
M.~J. Abraham, T.~Murtola, R.~Schulz, S.~P{\'a}ll, J.~C. Smith, B.~Hess, and
  E.~Lindahl, ``Gromacs: High performance molecular simulations through
  multi-level parallelism from laptops to supercomputers,'' {\em SoftwareX},
  vol.~1, pp.~19--25, 2015.

\bibitem{steinbach1994new}
P.~J. Steinbach and B.~R. Brooks, ``New spherical-cutoff methods for long-range
  forces in macromolecular simulation,'' {\em Journal of computational
  chemistry}, vol.~15, no.~7, pp.~667--683, 1994.

\bibitem{mackerell1998all}
A.~D. MacKerell~Jr, D.~Bashford, M.~Bellott, R.~L. Dunbrack~Jr, J.~D. Evanseck,
  M.~J. Field, S.~Fischer, J.~Gao, H.~Guo, S.~Ha, {\em et~al.}, ``All-atom
  empirical potential for molecular modeling and dynamics studies of
  proteins,'' {\em The journal of physical chemistry B}, vol.~102, no.~18,
  pp.~3586--3616, 1998.

\bibitem{ghorbani2021exploring}
M.~Ghorbani, B.~R. Brooks, and J.~B. Klauda, ``Exploring dynamics and network
  analysis of spike glycoprotein of sars-cov-2,'' {\em Biophysical Journal},
  vol.~120, no.~14, pp.~2902--2913, 2021.

\bibitem{khan2021preliminary}
A.~Khan, D.-Q. Wei, K.~Kousar, J.~Abubaker, S.~Ahmad, J.~Ali, F.~Al-Mulla,
  S.~S. Ali, N.~Nizam-Uddin, A.~Mohammad~Sayaf, {\em et~al.}, ``Preliminary
  structural data revealed that the sars-cov-2 b. 1.617 variant's rbd binds to
  ace2 receptor stronger than the wild type to enhance the infectivity,'' {\em
  ChemBioChem}, vol.~22, no.~16, pp.~2641--2649, 2021.

\bibitem{raghuvamsi2021sars}
P.~V. Raghuvamsi, N.~K. Tulsian, F.~Samsudin, X.~Qian, K.~Purushotorman,
  G.~Yue, M.~M. Kozma, W.~Y. Hwa, J.~Lescar, P.~J. Bond, {\em et~al.},
  ``Sars-cov-2 s protein: Ace2 interaction reveals novel allosteric targets,''
  {\em Elife}, vol.~10, p.~e63646, 2021.

\bibitem{frishman1995knowledge}
D.~Frishman and P.~Argos, ``Knowledge-based protein secondary structure
  assignment,'' {\em Proteins: Structure, Function, and Bioinformatics},
  vol.~23, no.~4, pp.~566--579, 1995.

\bibitem{zhou2003free}
R.~Zhou, ``Free energy landscape of protein folding in water: explicit vs.
  implicit solvent,'' {\em Proteins: Structure, Function, and Bioinformatics},
  vol.~53, no.~2, pp.~148--161, 2003.

\bibitem{calimet2001protein}
N.~Calimet, M.~Schaefer, and T.~Simonson, ``Protein molecular dynamics with the
  generalized born/ace solvent model,'' {\em Proteins: Structure, Function, and
  Bioinformatics}, vol.~45, no.~2, pp.~144--158, 2001.

\bibitem{verkhivker2020molecular}
G.~M. Verkhivker, ``Molecular simulations and network modeling reveal an
  allosteric signaling in the sars-cov-2 spike proteins,'' {\em Journal of
  proteome research}, vol.~19, no.~11, pp.~4587--4608, 2020.

\bibitem{singh2021serine}
A.~Singh, G.~Steinkellner, K.~K{\"o}chl, K.~Gruber, and C.~C. Gruber, ``Serine
  477 plays a crucial role in the interaction of the sars-cov-2 spike protein
  with the human receptor ace2,'' {\em Scientific reports}, vol.~11, no.~1,
  pp.~1--11, 2021.

\bibitem{ngo2021identifying}
V.~A. Ngo and R.~K. Jha, ``Identifying key determinants and dynamics of
  sars-cov-2/ace2 tight interaction,'' {\em PloS one}, vol.~16, no.~9,
  p.~e0257905, 2021.

\bibitem{chen2021machine}
Y.~Chen, A.~Kr{\"a}mer, N.~E. Charron, B.~E. Husic, C.~Clementi, and
  F.~No{\'e}, ``Machine learning implicit solvation for molecular dynamics,''
  {\em The Journal of Chemical Physics}, vol.~155, no.~8, p.~084101, 2021.

\bibitem{kulke2018replica}
M.~Kulke, N.~Geist, D.~\"Moller, and W.~Langel, ``Replica-based protein
  structure sampling methods: compromising between explicit and implicit
  solvents,'' {\em The Journal of Physical Chemistry B}, vol.~122, no.~29,
  pp.~7295--7307, 2018.

\bibitem{shao2018assessing}
Q.~Shao and W.~Zhu, ``Assessing amber force fields for protein folding in an
  implicit solvent,'' {\em Physical Chemistry Chemical Physics}, vol.~20,
  no.~10, pp.~7206--7216, 2018.

\bibitem{suenaga2003replica}
A.~Suenaga, ``Replica-exchange molecular dynamics simulations for a small-sized
  protein folding with implicit solvent,'' {\em Journal of Molecular Structure:
  THEOCHEM}, vol.~634, no.~1-3, pp.~235--241, 2003.

\bibitem{onufriev2018water}
A.~V. Onufriev and S.~Izadi, ``Water models for biomolecular simulations,''
  {\em Wiley Interdisciplinary Reviews: Computational Molecular Science},
  vol.~8, no.~2, p.~e1347, 2018.

\bibitem{roe2007secondary}
D.~R. Roe, A.~Okur, L.~Wickstrom, V.~Hornak, and C.~Simmerling, ``Secondary
  structure bias in generalized born solvent models: comparison of
  conformational ensembles and free energy of solvent polarization from
  explicit and implicit solvation,'' {\em The Journal of Physical Chemistry B},
  vol.~111, no.~7, pp.~1846--1857, 2007.

\bibitem{okur2006improved}
A.~Okur, L.~Wickstrom, M.~Layten, R.~Geney, K.~Song, V.~Hornak, and
  C.~Simmerling, ``Improved efficiency of replica exchange simulations through
  use of a hybrid explicit/implicit solvation model,'' {\em Journal of Chemical
  Theory and Computation}, vol.~2, no.~2, pp.~420--433, 2006.

\bibitem{deng2004hydration}
Y.~Deng and B.~Roux, ``Hydration of amino acid side chains: Nonpolar and
  electrostatic contributions calculated from staged molecular dynamics free
  energy simulations with explicit water molecules,'' {\em The Journal of
  Physical Chemistry B}, vol.~108, no.~42, pp.~16567--16576, 2004.

\bibitem{kumar2021differential}
V.~G. Kumar, D.~S. Ogden, U.~H. Isu, A.~Polasa, J.~Losey, and M.~Moradi,
  ``Differential dynamic behavior of prefusion spike proteins of sars
  coronaviruses 1 and 2,'' {\em BioRxiv}, pp.~2020--12, 2021.

\bibitem{piplani2021silico}
S.~Piplani, P.~K. Singh, D.~A. Winkler, and N.~Petrovsky, ``In silico
  comparison of sars-cov-2 spike protein-ace2 binding affinities across species
  and implications for virus origin,'' {\em Scientific reports}, vol.~11,
  no.~1, pp.~1--13, 2021.

\bibitem{tao2020using}
P.~Tao and Y.~Xiao, ``Using the generalized born surface area model to fold
  proteins yields more effective sampling while qualitatively preserving the
  folding landscape,'' {\em Physical Review E}, vol.~101, no.~6, p.~062417,
  2020.

\bibitem{gao2020short}
A.~Gao, R.~C. Remsing, and J.~D. Weeks, ``Short solvent model for ion
  correlations and hydrophobic association,'' {\em Proceedings of the National
  Academy of Sciences}, vol.~117, no.~3, pp.~1293--1302, 2020.

\bibitem{chakravorty2018reproducing}
A.~Chakravorty, Z.~Jia, L.~Li, S.~Zhao, and E.~Alexov, ``Reproducing the
  ensemble average polar solvation energy of a protein from a single structure:
  Gaussian-based smooth dielectric function for macromolecular modeling,'' {\em
  Journal of chemical theory and computation}, vol.~14, no.~2, pp.~1020--1032,
  2018.

\bibitem{stokely2010effect}
K.~Stokely, M.~G. Mazza, H.~E. Stanley, and G.~Franzese, ``Effect of hydrogen
  bond cooperativity on the behavior of water,'' {\em Proceedings of the
  National Academy of Sciences}, vol.~107, no.~4, pp.~1301--1306, 2010.

\bibitem{franzese2002liquid}
G.~Franzese and H.~E. Stanley, ``Liquid-liquid critical point in a hamiltonian
  model for water: analytic solution,'' {\em Journal of Physics: Condensed
  Matter}, vol.~14, no.~9, p.~2201, 2002.

\bibitem{ben2008one}
A.~Ben-Naim, ``One-dimensional model for water and aqueous solutions. i. pure
  liquid water,'' {\em The Journal of chemical physics}, vol.~128, no.~2,
  p.~024505, 2008.

\bibitem{lovett1969one}
R.~A. Lovett and A.~Ben-Naim, ``One-dimensional model for aqueous solutions of
  inert gases,'' {\em The Journal of Chemical Physics}, vol.~51, no.~7,
  pp.~3108--3119, 1969.

\bibitem{bianco2017stability}
V.~Bianco, N.~Pag{\`e}s-Gelabert, I.~Coluzza, and G.~Franzese, ``How the
  stability of a folded protein depends on interfacial water properties and
  residue-residue interactions,'' {\em Journal of Molecular Liquids}, vol.~245,
  pp.~129--139, 2017.

\bibitem{bianco2017role}
V.~Bianco, G.~Franzese, C.~Dellago, and I.~Coluzza, ``Role of water in the
  selection of stable proteins at ambient and extreme thermodynamic
  conditions,'' {\em Physical Review X}, vol.~7, no.~2, p.~021047, 2017.

\bibitem{ben1971statistical}
A.~Ben-Naim, ``Statistical mechanics of "waterlike" particles in two
  dimensions. i. physical model and application of the percus--yevick
  equation,'' {\em The Journal of Chemical Physics}, vol.~54, no.~9,
  pp.~3682--3695, 1971.

\bibitem{succi2014lattice}
S.~Succi, N.~Moradi, A.~Greiner, and S.~Melchionna, ``Lattice boltzmann
  modeling of water-like fluids,'' {\em Frontiers in Physics}, vol.~2, p.~22,
  2014.

\bibitem{barroso2019opep6}
F.~L. Barroso~da Silva, F.~Sterpone, and P.~Derreumaux, ``Opep6: A new
  constant-ph molecular dynamics simulation scheme with opep coarse-grained
  force field,'' {\em Journal of Chemical Theory and Computation}, vol.~15,
  no.~6, pp.~3875--3888, 2019.

\bibitem{arnarez2015dry}
C.~Arnarez, J.~J. Uusitalo, M.~F. Masman, H.~I. Ing{\'o}lfsson, D.~H. De~Jong,
  M.~N. Melo, X.~Periole, A.~H. De~Vries, and S.~J. Marrink, ``Dry martini, a
  coarse-grained force field for lipid membrane simulations with implicit
  solvent,'' {\em Journal of chemical theory and computation}, vol.~11, no.~1,
  pp.~260--275, 2015.

\bibitem{sterpone2018molecular}
F.~Sterpone, P.~Derreumaux, and S.~Melchionna, ``Molecular mechanism of protein
  unfolding under shear: A lattice boltzmann molecular dynamics study,'' {\em
  The Journal of Physical Chemistry B}, vol.~122, no.~5, pp.~1573--1579, 2018.

\end{thebibliography}

\end{document}